%% file: icml_ss_paper_cr.tex
\documentclass{article}

\usepackage{microtype}
\usepackage{graphicx}
\usepackage{subcaption}
\usepackage{booktabs}
\usepackage{hyperref}

\usepackage[preprint]{icml2026}
\makeatletter
\renewcommand{\Notice@String}{}
\makeatother

\usepackage{amsmath}
\usepackage{amssymb}
\usepackage{mathtools}
\usepackage{amsthm}
\usepackage[capitalize,noabbrev]{cleveref}
\usepackage{multirow}

\theoremstyle{plain}

\theoremstyle{definition}

\theoremstyle{remark}

\usepackage[textsize=tiny]{todonotes}

\icmltitlerunning{Flow Matching-Based Speech Source Separation with Best-of-N Biometric Sampling}

\begin{document}







\twocolumn[
  \icmltitle{Flow Matching-Based Speech Source Separation with Best-of-N Biometric Sampling}

  \begin{icmlauthorlist}
    \icmlauthor{Anastasia Zorkina}{itmo}
    \icmlauthor{Alexandr Anikin}{itmo}
    \icmlauthor{Nikita Khmelev}{itmo,stc}
    \icmlauthor{Anastasiya Korenevskaya}{itmo}
    \icmlauthor{Sergey Novoselov}{itmo,stc}
    \icmlauthor{Vladimir Volokhov}{itmo,stc}
    \icmlauthor{Maxim Korenevsky}{stc}
    \icmlauthor{Yuriy Matveev}{itmo}
  \end{icmlauthorlist}

  \icmlaffiliation{itmo}{ITMO University, Speech Processing Group, Russia}
  \icmlaffiliation{stc}{Speech Technology Center Ltd., R\&D department, Russia}

  \icmlcorrespondingauthor{Anastasia Zorkina}{zorkina@speechpro.com}

  \icmlkeywords{source separation, flow matching, speech enhancement, generative models, best-of-N sampling}

  \vskip 0.3in
]

\printAffiliationsAndNotice{}

\begin{abstract}
Single-channel speech separation remains challenging for real-world deployment due to source permutation ambiguity, sampling variability of generative models, and the difficulty of processing long recordings with chunk-wise inference. We address these issues with a conditional flow-matching-based method that produces an ordered two-source output conditioned on the mixture. A frozen speaker encoder defines the source order during training and is reused at inference for biometric best-of-$N$ candidate selection and chunk-level channel alignment. We evaluate separation quality on Libri2Mix benchmark using SI-SDR, PESQ, and ESTOI, and measure downstream impact using cpWER for automatic speech recognition and EER for speaker verification. The results show that the proposed Transformer U-Net variant is competitive with strong baselines in objective separation metrics and achieves the lowest downstream automatic speech recognition and speaker verification error rates in all evaluated settings.


\end{abstract}

\section{Introduction}

Speech source separation, also known as the ``cocktail party problem'', is a fundamental task in audio processing and speech technologies \cite{li2025advances,araki2025thirty,wang2025speechsurvey}. It aims to extract individual speech signals from a mixture, benefiting systems such as automatic speech recognition (ASR), speaker recognition (SR), intelligent voice assistants, and others. Despite rapid progress driven by deep learning, speech separation remains challenging due to acoustic complexity, the ill-posed nature of single-channel inversion, varying overlap patterns, permutation ambiguity, and high speaker variability.

Modern deterministic speech source separation systems based on Transformer and Conformer architectures achieve SI-SDRi above 24 dB on the WSJ0-2mix benchmark \cite{zhao2024mossformer2,shin2024separate}. Generative counterparts, including hybrid schemes \cite{lutati2023separate,wang2024fastgenerative} and fully generative diffusion or flow matching models \cite{scheibler2023diffusion,dong2025edsep, scheibler2025floss}, offer new modeling paradigms that improve perceptual naturalness at the cost of sampling variance and higher computation.

Deploying these systems in practice faces additional hurdles: non-causality, processing of long recordings, downstream integration, and computational efficiency. In this work, we propose a practical speech separation system based on conditional flow matching, built on generative speech enhancement models from the NVIDIA NeMo Toolkit \cite{jukic2024schrodinger,ku2025generative}. We formulate two-speaker separation as a conditional generation of a structured waveform that contains both separated sources. Besides, we introduce a best-of-$N$ biometric sampling procedure that selects the most speaker-disentangled generation among multiple stochastic candidates.

The main contributions of this work are as follows: 1) A flow-matching-based speech separation method adapted from generative speech enhancement. 2) A best-of-$N$ biometric criterion for inference-time candidate selection. 3)~Long-form processing via chunk-wise generation combined with speaker-recognition-based channel tracking. 4) Evaluation of downstream ASR and SR performance after use of our speech source separation system.

\section{Background}
\label{sec:background}

\paragraph{Speech source separation.}
Formally, the task is to recover $K$ signals $s_1,\ldots,s_K$ from a mixture $m=\sum_k s_k$. Permutation-invariant training (PIT)~\cite{yu2017pit,kolbaek2017upit} resolves source-ordering symmetry at the loss level but leaves output channels interchangeable at inference, complicating block-wise processing of long-form audio. An information-theoretic upper bound of $23.1$\,dB SI-SDRi has been established for deterministic separators~\cite{lutati2022sepit}; under heavy overlap, they also introduce perceptual artefacts that SI-SDR does not capture but that downstream ASR/SR penalise~\cite{wang2025speechsurvey}. Generative separators sidestep both issues by sampling from $p(s_1,\ldots,s_K\mid m)$ via diffusion or conditional flow matching~\cite{lipman2023flow}, at the price of run-to-run sampling variance for which no principled inference-time selection criterion has been proposed.

\paragraph{Speaker recognition.}
Speaker recognition is the task of identifying a speaker from an audio recording. Its canonical formulation is speaker verification (SV): determining whether two recordings originate from the same speaker. State-of-the-art systems are based on deep neural networks, either trained from scratch for this task (e.g., ECAPA-TDNN~\cite{desplanques2020ecapa}, ReDimNet~\cite{yakovlev2024reshape}) or adapted from speech foundation models (e.g., Wav2Vec~2.0-TDNN~\cite{novoselov2022robust}). Training is performed as a speaker classification problem (7,205 classes on VoxCeleb 1 \& 2~\cite{nagrani2017voxceleb}), employing feature normalization and an additive angular margin in the softmax objective (AAM-Softmax) to enhance class separability. At inference time, the classification layer is discarded, and the model extracts a compact speaker embedding; embeddings from different recordings are then compared via cosine similarity.

\paragraph{Best-of-$\boldsymbol{N}$ sampling.}
Best-of-$N$ returns the highest-ranked of $N$ independent generator samples under an external verifier, trading inference compute for quality without retraining the generator. It underpins LLM alignment \cite{stiennon2020bon}, verifier-based mathematical reasoning \cite{cobbe2021verifiers}, code generation \cite{chen2021codex}, and inference-time scaling of text-to-image diffusion \cite{xie2025sana15,choi2025plateaus}.

\section{Proposed Method}
\label{sec:method}

\subsection{Conditional Flow Matching Formulation}

Flow matching defines generation through a time-dependent vector field whose flow map transports samples from a simple prior distribution to the data distribution. Let $\phi_t$ denote the flow map induced by a vector field $u_t$,
\begin{equation}
    \frac{d}{dt}\phi_t(x) = u_t(\phi_t(x)),
    \qquad \phi_0(x)=x .
\end{equation}
In conditional generation, we approximate this vector field with a neural estimator $v_t(x_t,c;\theta)$, where $x_t$ is the current state, $c$ is the conditioning information, and $\theta$ are trainable parameters.

To obtain a tractable training target, we follow the optimal-transport conditional path~\cite{lipman2023flow}. Given $x_1 \sim p_{\mathrm{data}}(\cdot \mid c)$, $x_0 \sim \mathcal{N}(0,I)$, $t \sim \mathcal{U}[0,1]$, and an intermediate state $x_t$ sampled from this path, the estimator is trained with
\begin{equation}
    \mathcal{L}_{\mathrm{CFM}}(\theta)
    =
    \mathbb{E}
    \left[
    \left\|
    v_t(x_t,c;\theta)
    -
    u_t(x_t \mid x_0,x_1)
    \right\|_2^2
    \right].
\end{equation}

We formulate two-speaker speech separation as a conditional generation, guided by some feature $c$. The observed mixture waveform is $m=s_1+s_2$, where $s_1$ and $s_2$ are speech signals from two speakers. For each mixture, $c$ is obtained from the complex STFT features of the mixture-derived signal, while $x_1$ denotes the complex STFT features of the corresponding separated-source signal; therefore, both predicted and target velocities are defined in the same STFT feature space.

\subsection{Training Procedure}

\begin{figure}
    \centering
    \includegraphics[width=1\linewidth]{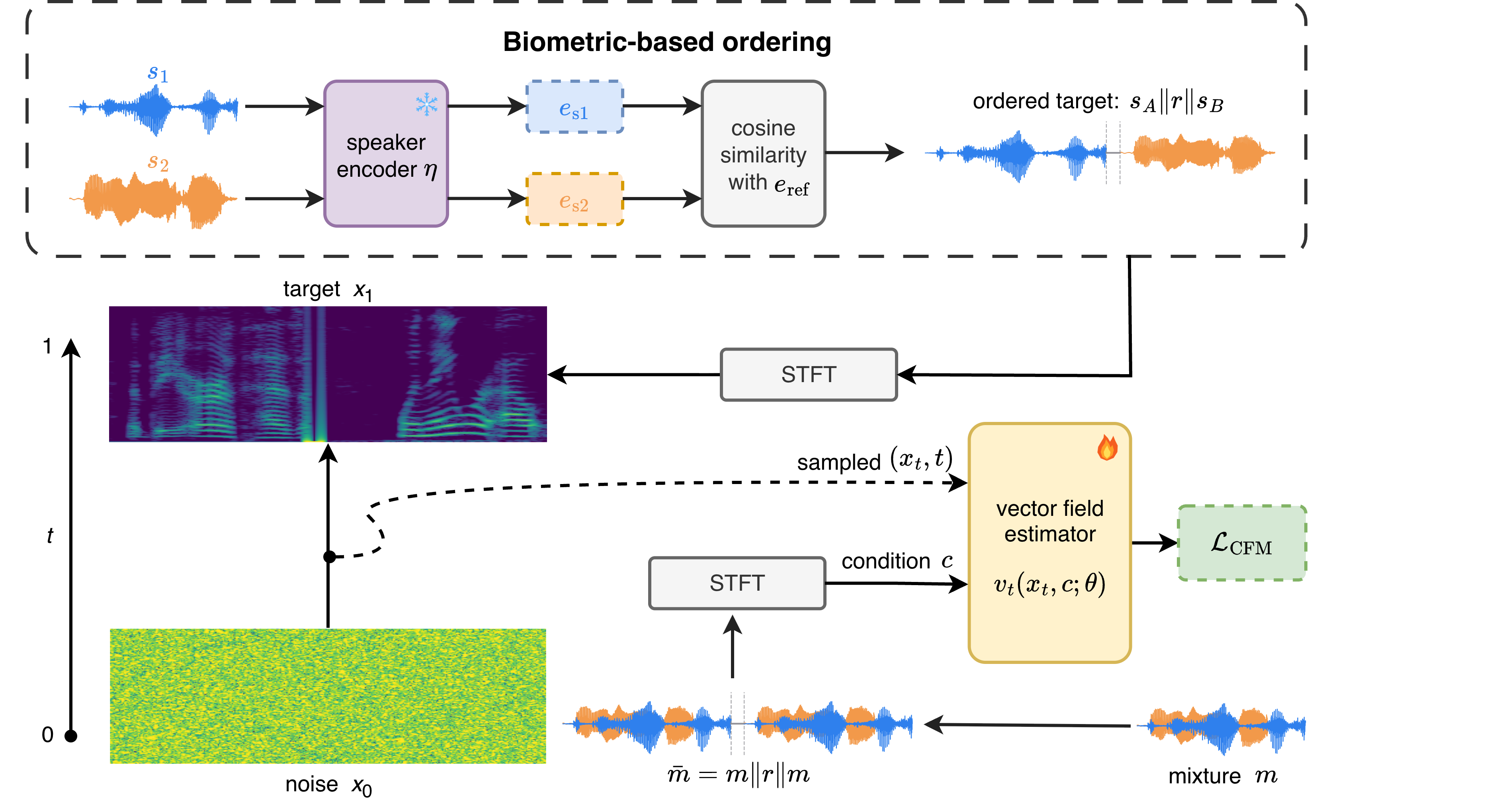}
    \caption{Training procedure of the proposed conditional flow matching demixer. }
    \label{fig:training}
\end{figure}

The training pipeline is shown in Fig.~\ref{fig:training}. Each training example consists of a mixture $m$ and two clean sources $(s_1,s_2)$; however, these references do not by themselves define a canonical target order. Single-channel separation is ill-posed and permutation ambiguous: the source order and its swapped version define the same mixture, while the model must learn a deterministic output order. To make the target closer to well-defined, we resolve this ambiguity using biometric comparison with a pretrained speaker encoder $\eta$: source embeddings are compared with a predefined reference speaker embedding $e_{\mathrm{ref}}$ using cosine similarity, yielding an ordered pair $(s_A,s_B)$, where $s_A$ is the source closer to $e_{\mathrm{ref}}$. We concatenate the ordered sources through a fixed non-speech separator, $\bar{x}=s_A \Vert r \Vert s_B$, and duplicate the mixture in the same layout, $\bar{m}=m \Vert r \Vert m$; their complex STFT features define $x_1$ and $c$, respectively. Thus, the source separation task is reduced to a denoising task, where the structural interference of the reversely oriented speech signal $s_B \Vert r \Vert s_A$ should be removed from $\bar{m}$.

During training, we sample $x_0 \sim \mathcal{N}(0,I)$ and $t \sim \mathcal{U}(0,1)$, then form $x_t$ using the conditional path above. The neural estimator receives $x_t$, the conditioning representation $c$, and the time $t$, and is optimized with the conditional flow matching loss. Conditional dropout is applied by replacing $c$ with zeros with probability $1-p_{\mathrm{cond}}$, exposing the estimator to both conditional and unconditional inputs.

\subsection{Inference Pipeline}

Conventional source separation systems typically operate on a complete input segment, which limits their applicability to long-form recordings. Our inference pipeline instead targets recordings of arbitrary duration by processing the mixture chunk by chunk. This requires two stages: local generation for each chunk and subsequent alignment of chunk-level outputs.

\paragraph{Chunk-level generation.}
Given a mixture recording $m$, we split it into 1-second chunks with a 0.5-second hop. Each chunk $m_c$ is processed independently using the same representation as in training: the mixture-derived condition is encoded into complex STFT features, the initial state is sampled as Gaussian noise $x_0 \sim \mathcal{N}(0,I)$, and the learned ODE is solved with an Euler sampler to produce a local source pair $(\hat{s}_{c,A},\hat{s}_{c,B})$.

The stochasticity of flow-based generation allows us to draw multiple plausible separations for the same chunk. We exploit this at inference time through best-of-$N$ biometric sampling: from $N$ independently generated candidates, we select the pair whose output-channel speaker embeddings are least similar,
\begin{equation}
    i^*
    =
    \arg\min_i
    \operatorname{cos}
    \left(
    \eta(\hat{s}_{c,A}^{(i)}),
    \eta(\hat{s}_{c,B}^{(i)})
    \right).
\end{equation}
This favors candidates in which the two generated channels correspond to different speakers.

\paragraph{Chunk alignment.}
After chunk-level generation, the local channel order must be made consistent before reconstruction, since the two generated sources may swap between adjacent chunks. We use biometric alignment based on the speaker embeddings of the generated channels. For each global output channel, embeddings from accepted chunks are accumulated and clustered into two centroids, which gives a more robust channel representation in the presence of noise, leakage, or locally unreliable generations. A new chunk is assigned to the channel order whose embeddings best match the current centroids; otherwise, its two channels are swapped. For controlled analysis, we also report an oracle SI-SDR alignment, which chooses the chunk permutation using ground-truth sources and serves as a reference alignment upper bound. Finally, the aligned chunks are placed back at their original time positions and combined by overlap-add averaging, yielding two full-length separated waveforms.

\section{Experimental Setup}
\label{sec:experiments}

\subsection{Data and Preprocessing}

We use the two-speaker Libri2Mix benchmark from the LibriMix corpus~\cite{cosentino2020librimix}: 16~kHz max, train-360 for training, and the corresponding test subsets for evaluation. Experiments are reported for the official mix\_clean and mix\_both mixture types.

We additionally apply online augmentation. We load 2-second synchronized mixture/source segments, reverberate both sources with independently selected speaker RIRs, mix them with target-to-interferer SNR sampled from $[0,5]$~dB, and add WHAM!~\cite{wichern2019wham}, MUSAN~\cite{snyder2015musan}, or CHiME-8~\cite{cornell2024chime8} noise to the mixture with SNR sampled from $[10,20]$~dB. Targets are the reverberated clean sources. We keep the central 1-second region after augmentation to suppress convolution boundary artifacts.

Targets are represented as $s_A \Vert r \Vert s_B$ and conditions as $m \Vert r \Vert m$, where $r$ is a 1000-sample constant separator with amplitude $0.5$ (62.5~ms at 16~kHz). Following the NVIDIA NeMo Toolkit implementation~\cite{kuchaiev2019nemo}, we use complex STFT features with hop length 128, magnitude power 0.5, and scale factor 0.33. The FFT length is 254 for NCSN++ (ConvUnet) and 510 for Transformer U-Net (TUnet).

\subsection{Models and Baselines}

We evaluate two velocity estimators: an NCSN++ spectrogram network~\cite{song2021scorebased}, referred to as ConvUnet in the results, and a 24-layer Spectrogram Transformer U-Net (TUnet) initialized from the NVIDIA generative speech restoration checkpoint~\cite{ku2025generative}. Source ordering, best-of-$N$ selection, and chunk alignment use a frozen Wav2Vec~2.0-based speaker embedding model~\cite{novoselov2022robust}.

Training uses Adam optimizer, learning rate $10^{-4}$ with cosine decay to $10^{-6}$, gradient clipping, and EMA decay $0.999$. NCSN++ is trained for up to 300 epochs on NVIDIA A100 GPUs; Transformer U-Net is trained for up to 150 epochs on one A100. We use conditional dropout with $p_{\mathrm{cond}}=0.99$.

Baselines are DiffSep~\cite{scheibler2023diffusion},  and SepReformer~\cite{shin2024separate} (Large model evaluated in both full-utterance and chunked modes). Additionally recent target speaker extraction method MeanFlow-TSE~\cite{shimizu2025meanflowtse} was evaluated and compared.For downstream evaluation, we measure automatic speech recognition performance with Whisper V3~\cite{radford2023robust} and speaker recognition performance with Wav2Vec~2.0~\cite{khmelev2025optimal} system.

\subsection{Evaluation Protocols and Metrics}

Separation quality is measured in terms of SI-SDR, PESQ, and ESTOI. Downstream ASR quality is measured in terms of cpWER and 
SV quality is measured in terms of EER. For EER evaluation, we generate verification protocols with clean source utterances as enrollment samples and separated outputs as test samples. For the proposed method all metrics are computed after best-of-$N$ selection with $N=4$. 

\input{tables/objective-metrics}
\section{Results}
\label{sec:results}

Table~\ref{tab:table_1} reports objective separation metrics. In the ``clean'' setting, TUnet achieves the best ESTOI (0.93) and competitive SI-SDR (17.30 dB), substantially outperforming DiffSep (9.60 dB SI-SDR). SepReformer (full) obtains the highest SI-SDR (19.22 dB), but this configuration processes the full utterance at once, which makes it unsuitable for long recordings in real-world scenarios. In the chunked setting, SepReformer drops to 11.30 dB SI-SDR and 0.88 ESTOI. MeanFlow-TSE achieves the best PESQ (3.27) and the best SI-SDR/PESQ on the ``both'' subset, but it addresses target speaker extraction and uses target-speaker reference information, making it a different problem setting from blind two-speaker separation.  Overall, the proposed ConvUnet and TUnet consistently outperform DiffSep and remain competitive with state-of-the-art approaches across both conditions.

\begin{figure}
    \centering
    \includegraphics[width=1\linewidth]{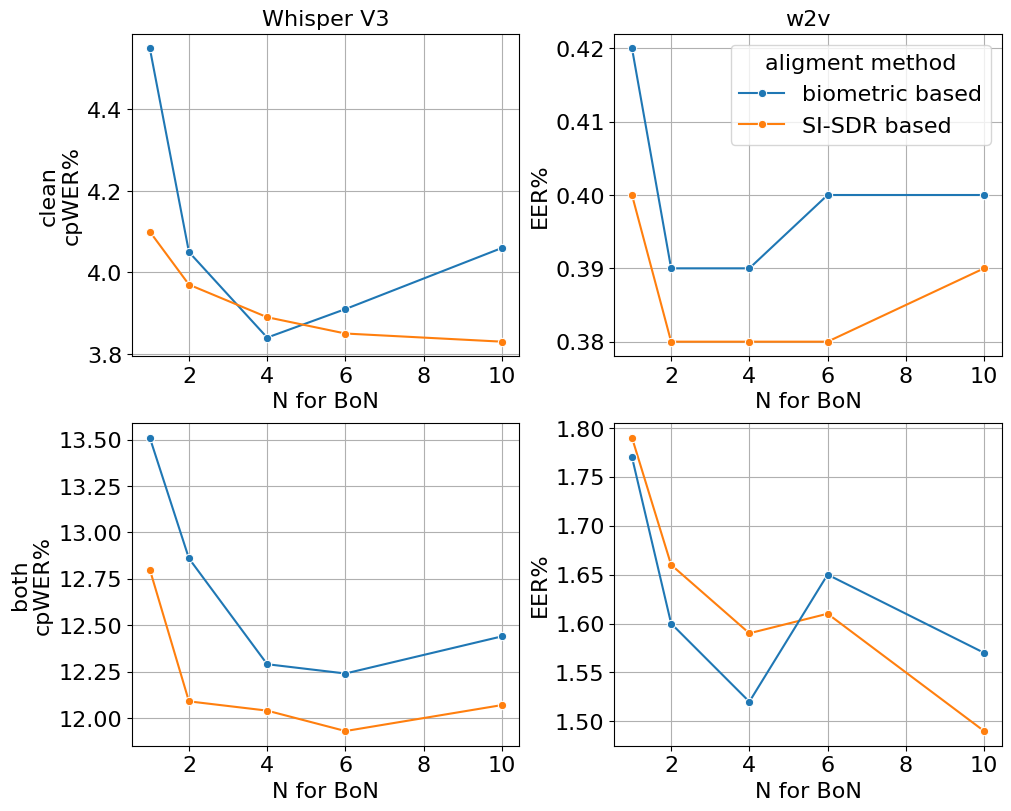}
    \caption{Best-of-$N$ selection on the Libri2Mix test set. Top: ``clean''; bottom: ``both''. Left: cpWER (\%) with Whisper~V3 ASR; right: EER (\%) with Wav2Vec~2.0 speaker verification. Blue: proposed biometric selection; red: SI-SDR oracle using ground-truth references.}
    \label{fig:bon-selection}
\end{figure}

\Cref{fig:bon-selection} analyzes best-of-$N$ sampling for the proposed TUnet model. Additionally it compares biometric channel selection pipeline with SI-SDR-based oracle selection using ground-truth sources. 
Increasing N generally reduces both cpWER and EER for both strategies, with saturation starting around $N=4$. The biometric-based criterion, which does not require ground-truth references, closely approaches the oracle performance, demonstrating its practical utility for inference-time candidate selection in generative separation.

\Cref{tab:table_2} reports downstream ASR and SV performance after speech separation on the Libri2Mix clean test set. TUnet achieves the best downstream results among all evaluated systems, with 3.84\% cpWER and 0.39\% EER. ConvUnet also remains competitive, improving over MeanFlow-TSE and SepReformer (chunk) on both metrics. Although SepReformer (full) performs close to ConvUnet in the full-utterance setting, its chunk-wise variant degrades substantially, which is consistent with the objective results in \Cref{tab:table_1}. An additional ablation with alternative downstream backends shows the same trend for Parakeet ASR~\citep{sekoyan2025canary} and for ResNet-34 and DistillWhisper-based~\citep{dwhisper2026asv} SV models, confirming that the observed gains are not specific to a single ASR or SV system. Overall, the downstream evaluation shows that the proposed TUnet is the strongest option in practical chunk-wise processing among the evaluated systems.

\input{tables/asr-asv_compact}

\section{Conclusion}
\label{sec:conclusion}
We presented a novel conditional flow matching source separation approach that combines biometric source ordering, best-of-$N$ candidate selection, and chunk-wise processing for long recordings. On Libri2Mix, the Transformer U-Net variant remains competitive in objective separation metrics and achieves the best downstream  cpWER and EER among the evaluated systems. These results indicate that the proposed system is well suited for practical speech processing pipelines with downstream ASR and SV tasks.

\bibliography{icml_ss_paper}
\bibliographystyle{icml2026}

\end{document}

%% file: tables/objective-metrics.tex

\begin{table}[t]
\centering
\caption{Performance comparison of speech source separation and target speaker extraction methods on the Libri2Mix test set in terms of SI-SDR (dB), PESQ, and ESTOI. Values are reported as ``clean'' / ``both''.}
\label{tab:table_1}

\resizebox{0.48\textwidth}{!}{
\begin{tabular}{|l|c|c|c|}
\hline

\textbf{Model} &
\textbf{SI-SDR} &
\textbf{PESQ} &
\textbf{ESTOI}
\\ \hline

ConvUnet (our)
& 14.42 / 9.28
& 2.64 / 1.65
& 0.90 / 0.75
\\ \hline

TUnet (our)
& 17.30 / 11.25
& 3.11 / 1.92
& \textbf{0.93} / 0.79
\\ \hline

DiffSep
& 9.60 / -
& 2.58 / -
& 0.78 / -
\\ \hline

SepReformer (chunk)
& 11.30 / -
& 2.45 / -
& 0.88 / -
\\ \hline

SepReformer (full)
& \textbf{19.22} / \textbf{13.70}
& 3.02 / 2.14
& 0.92 / \textbf{0.83}
\\ \hline

MeanFlow-TSE (full)
& 17.56 / 11.68
& \textbf{3.27} / \textbf{2.18}
& 0.91 / {0.80}
\\ \hline

\end{tabular}
}
\end{table}















%% file: tables/asr-asv_compact.tex










\begin{table}[t]
\centering
\caption{Downstream ASR and SV performance on Libri2Mix clean after speech separation. Metrics: cpWER (\%) and EER (\%).}
\label{tab:table_2}

\resizebox{0.48\textwidth}{!}{
\begin{tabular}{|l|c|c|}
\hline

\textbf{Model} &
\textbf{Whisper V3, cpWER} &
\textbf{Wav2Vec 2.0, EER}
\\ \hline

ConvUnet (our)
& 4.99
& 0.51
\\ \hline

TUnet (our)
& \textbf{3.84}
& \textbf{0.39}
\\ \hline

MeanFlow-TSE (full)
& 9.05
& 2.94
\\ \hline

SepReformer (full)
& 5.02
& 0.72
\\ \hline

SepReformer (chunk)
& 12.90
& 1.45
\\ \hline

\end{tabular}
}
\end{table}